\documentclass[12pt,twoside,fleqn]{article}
\usepackage{espcrc1}
\usepackage{epsfig}

 \newcommand{\AmS}{{\protect\the\textfont2
A\kern-.1667em\lower.5ex\hbox{M}\kern-.125emS}}

\title{$\pi$-$N$\/ charge exchange and $\pi^+$-$\pi^0$\/ scattering at low
       energies\thanks{Work supported by the U.S. National Science
       Foundation.}}

\author{D. Po\v{c}ani\'c and E. Frle\v{z}\\ \vskip 3mm Department of
Physics, University of Virginia, Charlottesville, VA 22901, U.S.A.}

\begin{document}
\maketitle

\begin{abstract}
$\pi$-$N$\/ and $\pi$-$\pi$\/ interactions near threshold are uniquely
sensitive to the chiral symmetry breaking part of the strong interaction.
The $\pi$-$N$\/ $\sigma$-term value with its implications for nucleon quark
structure and the recent controversy concerning the size of the scalar
quark condensate have renewed the experimental interest in these two
fundamental systems.  We report new differential cross sections for the
reaction $\pi^-p$$\to$$\pi^0n$\/ at 27.5 MeV pion incident kinetic energy,
measured between $\theta_{\rm CM}$ = 0$^\circ$ and 55$^\circ$.  Our results
are in excellent agreement with the existing comprehensive $\pi N$\/ phase
shift analysis.  We also report on a Chew-Low analysis of exclusive
$\pi^+p$$\to$$\pi^+\pi^0p$\/ data at 260 MeV pion incident energy.
\end{abstract}

\section{\boldmath $\pi$-$N$\/ CHARGE EXCHANGE AT 27.5 MeV}

While the basic mechanism of spontaneous breaking of chiral symmetry is
reasonably well in hand, certain aspects of the explicit breaking of chiral
symmetry ($\chi$SB), due to nonzero quark masses, remain not fully
resolved to date.  In the $\pi$-$N$\/ system at low energies the quantities
of interest are the chiral symmetry breaking ``sigma term'' and the
scattering lengths.  In particular, the $\sigma$-term has been found to
have an unexpectedly large value (for the most recent comprehensive
analysis see Ref.\ \cite{Gas-91}).  The discrepancy between the
$\sigma$-term values obtained from the baryon mass splitting and from
extrapolation of the isospin-even $\pi$-$N$\/ scattering amplitude has been
attributed to a nonzero $\bar{s}s$\/ content of the nucleon \cite{Gas-82}.
$\pi$-$N$\/ scattering lengths are related quantities that provide an
independent check of the chiral lagrangians.  Thus, low energy $\pi$-$N$\/
interactions have retained a fundamental significance and interest over the
years.

Unfortunately, inconsistencies in the existing $\pi$-$N$\/ data set have
given rise to significant uncertainties of the low energy $\pi$-$N$\/
amplitudes.  These, in turn, are reflected in the error limits of the
extracted ``experimental'' value of the $\sigma$-term.  This situation has
led to an effort to remeasure all low energy $\pi$-$N$\/ observables at the
remaining meson facilities.  In this work we focus on the charge exchange
reaction below 30 MeV pion incident energy.

Absolute measurements of the pion-nucleon charge exchange reaction
$\pi^-p$$\to$$\pi^0n$\/ below 50 or even 100 MeV are sparse.  The
difficulties stem from the requirement that the beam composition, beam
flux, and the $\pi^0$ detection efficiency all have to be measured or
determined accurately in an absolute way.

Early published data below 50 MeV were measured by detecting the neutron at
a single angle, 0$^\circ$, corresponding to the $\pi^0$ angle of
180$^\circ$ \cite{Duc-73}.  Another set of measurements \cite{Sal-84} used
a large NaI(Tl) crystal counter to detect single photons from the final
state $\pi^0$\/ decay, at 27.4 and 39.3 MeV incident pion energy, covering
a wide angular range.  Due to the nature of this method, yields from a
broad range of $\pi^0$ angles were mixed in at any given laboratory angle
of a detected single photon.  Hence, the authors could only report a
Legendre polynomial decomposition of the $\pi^-p$$\to$$\pi^0n$\/ angular
distribution, up to order 2.

First published direct measurements of angular distribution data of the
$\pi^-p$$\to$$\pi^0n$\/ reaction below 50 MeV were made using the LAMPF
$\pi^0$ spectrometer \cite{Bae-81} at seven energies between 32.5 and 63.5
MeV for $\theta_{\rm lab}(\pi^0) = 0 - 30^\circ$ \cite{Fit-86}.  A device
such as the $\pi^0$ spectrometer detects the two photons following the
$\pi^0$\/ decay in coincidence.  This, in turn, enables a full
reconstruction of the neutral pion's momentum four-vector.  In our work we
used the same technique.

\subsection{Experimental method and normalization}

The present measurements of the differential cross sections for the
reaction $\pi^-p$$\to$$\pi^0n$\/ at $27.5 \pm 0.2$ MeV were carried out in
the LEP secondary beam channel at the Clinton P. Anderson Meson
Physics Facility (LAMPF) in Los Alamos.  We used a weakly focusing 30 MeV
$\pi^-$\/ beam tune with 12 mr divergence (both horizontal and vertical), a
beam spot size of 9 mm FWHM, momentum spread $\Delta p/p = 3$\% and pion
flux averaging $5\times 10^5$\ $\pi^-$/sec.

Relative on-target beam intensity was monitored with a sealed ion gas
chamber in combination with a high precision charge integrator.  Absolute cross-calibration of chamber
ionization counts was obtained through activation measurements of the
$^{12}$C($\pi^-,\pi^-n)^{11}$C reaction using cylindrical plastic
scintillator targets \cite{But-82}.  The $^{11}$C activation measurements
were reproducible to better than $\pm 2$\%, while the $^{11}$C activation
cross section used for normalization has an uncertainty of 4.7\%
\cite{Lei-90}.

The electron and muon contaminations in the beam were determined by a
combination of direct measurement and constraints using the integrated
energy deposited in the sealed ion chamber that was calibrated in absolute
terms independently.  The associated uncertainty of the pion flux amounted
to $\pm 2.4$\%.

Our measurements were carried out using a $711 \pm 2$ mg/cm$^2$
polyethylene (CH$_2$) target, with a suitable $^{12}$C target for
background subtraction.  In addition, we recorded charge exchange data
using a $267 \pm 7$ mg/cm$^2$ liquid hydrogen target as a check.

We used the LAMPF $\pi^0$ spectrometer to detect coincident photons
following $\pi^0$ decay in $\pi^-p$$\to$$\pi^0n$.  The spectrometer
multiwire proportional chamber and veto counter efficiencies were
calibrated independently using cosmic muons.  All tracking efficiencies
were also evaluated from data and compared with a detailed simulation using
{\tt GEANT} \cite{Bru-87}. The resulting uncertainty of the integral
$\pi^0$ detection efficiency was 4.6\%.

The measured detector response to $\pi^0$'s from the charge exchange
reaction under study was compared to simulations using {\tt GEANT} and {\tt
PIANG} \cite{Gil-79}, with excellent agreement.  The rms angular resolution
of the spectrometer was 2$^\circ$.

\subsection{Results and discussion}

\begin{figure}[hbt]
\noindent \epsfig{figure=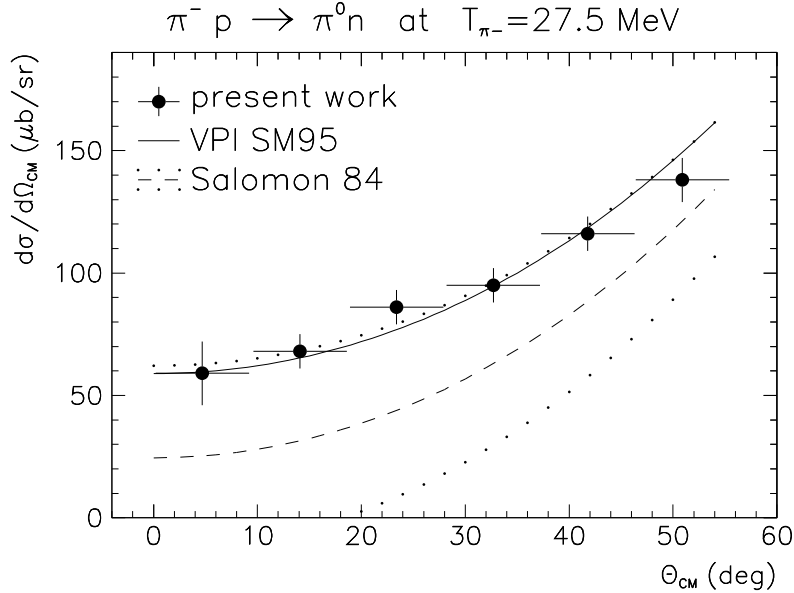,width=105mm} \hfill 
\begin{minipage}[b]{50mm}
\caption{Measured c.m.\ $\pi^-p$$\to$$\pi^0n$ differential cross sections
(full circles). Vertical bars reflect the statistical uncertainties only;
an overall normalization uncertainty of 7.5\% also applies.  Horizontal
bars denote the angular bin size; rms angular resolution was 2$^\circ$.
Solid curve: VPI SM95 partial wave analysis\ \protect\cite{said}.  Dashed
and dotted curves: Legendre polynomial fit to data from Ref.\
\protect\cite{Sal-84}.}  \vglue 2mm
\label{f:ang-dist}
\end{minipage}
\end{figure}

Results of our measurements of the $\pi^-p$$\to$$\pi^0n$ angular
distribution between 0$^\circ$ and 55$^\circ$ (c.m.), binned into $9^\circ$
wide bins, are plotted in Fig.~\ref{f:ang-dist} as full circles.  Error
bars shown in the figure reflect only statistical uncertainties; in
addition, an overall normalization uncertainty of 7.5\% applies to the
data, as discussed above.

For the sake of comparison we have also included in Fig.~\ref{f:ang-dist}
the angular distribution predicted by the comprehensive $\pi$-$N$ phase
shift analysis SM95 by the VPI group \cite{said} (solid curve).  The
agreement between our data and the VPI phase shift prediction is excellent.
Older data at this energy from Ref.\ \cite{Sal-84} are available only in
the form of a Legendre polynomial decomposition (fit) of the angular
distribution.  That fit is represented in Fig.\ \ref{f:ang-dist} by a
dashed line, while dotted lines denote the associated error limits.

\vglue -12mm
\vglue -5mm

In summary, our results provide a new stringent constraint on the low
energy $\pi$-$N$ phase shifts, and are in excellent agreement with the
existing body of $\pi$-$N$ data.

\section{Reaction \boldmath $\pi^+p\to\pi^+\pi^0p$ at 260 MeV}

Low energy $\pi$-$\pi$ scattering has enjoyed longstanding attention as a
window into the mechanism of chiral symmetry breaking.  Pion-pion
scattering lengths have recently come sharply into focus due to the
controversy regarding $\langle 0|\bar{q}q|0\rangle$, the scalar quark
condensate, and the two radically different and far-reaching scenarios of
$\chi$SB
\cite{Kne-95}.  The current most reliable value of $a_0^0 = 0.26 \pm 0.05\
\mu^{-1}$ (where $\mu \equiv m_\pi$), extracted mainly from $K_{e4}$ decay
data \cite{Poc-95}, is not accurate enough to make the required
distinction. 

We report here on preliminary results of a Chew-Low analysis\ \cite{Che-59}
of exclusive $\pi^+p \to \pi^+\pi^0p$ data measured at 260 MeV $\pi^+$
incident energy.  The experimental apparatus and the total cross section
analysis are described in Ref.\ \cite{Poc-94}.  The Chew-Low method
evaluates $\pi\pi$ cross sections by extrapolating to the pion pole the
function $F(s,t,m_{\pi\pi})$:
\begin{equation}
  \sigma_{\pi\pi}(m_{\pi\pi}) = \lim_{t\to\mu^2} F(s,t,m_{\pi\pi}) 
  = \lim_{t\to\mu^2} {\partial^2 \sigma_{\pi\pi N}(s) \over \partial t \,
    \partial m_{\pi\pi}} \cdot {\pi \over f_\pi^2} \cdot
    {p^2(t-\mu^2)^2 \over t\, m_{\pi\pi}(m_{\pi\pi}^2 - 4\mu^2)^{1/2}} ~,
  \label{e:chew-low}
\end{equation}
where $m_{\pi\pi}$\/ is the dipion invariant mass, $t$\/ is the squared
4-momentum transfer to the proton, $\sqrt{s}$\/ is the total c.m.\ energy,
$p$\/ is the incident pion momentum, and $f_\pi$ the pion decay constant.
In this work, we had to perform a deconvolution of the instrumental
resolution function from the data before we could construct an
interpretable $F(s,t,m_{\pi\pi})$.

\begin{figure}[htb]
\noindent \epsfig{figure=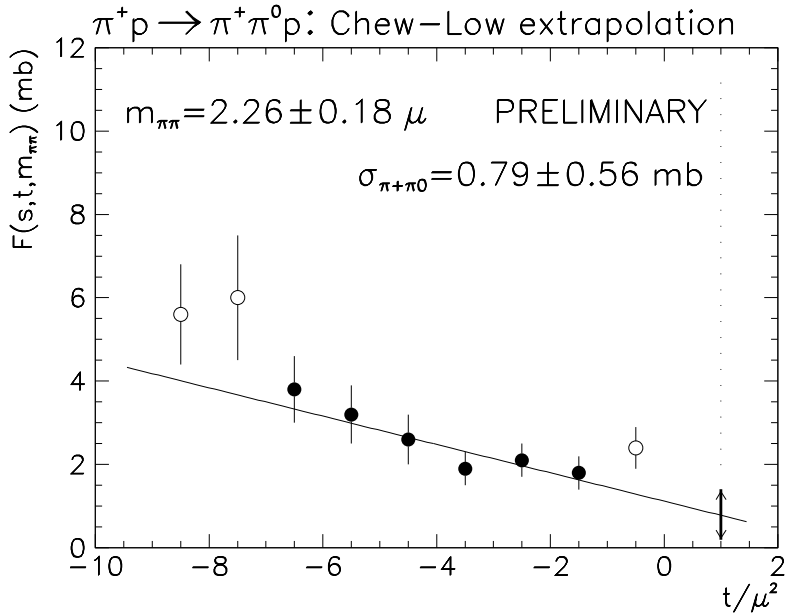,width=105mm} \hfill 
\begin{minipage}[b]{50mm}
\caption{Chew-Low function $F(s,t,m_{\pi\pi})$ con\-struc\-ted from $\pi^+p
\to \pi^+\pi^0p$ exclusive cross sections at 260 MeV is plotted as a function
of $t$ along with a linear fit (preliminary).  Full circles: data points
included in the fit.  Open circles: data points excluded from the fit.
The ex\-tra\-po\-la\-ted value of the $\pi\pi$ total cross section at
$m_{\pi\pi} = 2.26 \pm 0.18\ \mu$ is indicated.} \vglue 7mm 
\label{f:chew-low}
\end{minipage}
\end{figure}

Preliminary values of $F$, the Chew-Low extrapolation function, calculated
from our $\pi^+\pi^0p$ data at 260 MeV are plotted against $t$ in
Fig.~\ref{f:chew-low}, alongside a linear fit.  Points with $|t| > 7 \mu$
were excluded from the fit due to the diminishing contribution of the one
pion exchange process; the lowest $t$ point was excluded due to
deconvolution uncertainties.

\vglue -10mm
\vglue -5mm

Using the new $\pi^+\pi^0$ cross section datum we can deduce $a_0^2 \simeq
0.55 \pm 0.24\ \mu^{-1}$.  However, further work is required in order to
extract a more reliable extrapolated value of $\sigma(\pi\pi)$.  That
result, in turn, will be added to the existing $\pi\pi$ data set for a
comprehensive dispersion-relation analysis.

We gratefully acknowledge valuable contributions by S. Bruch and
R.C. Minehart in the analysis, and by the other members of the E1179
collaboration \cite{Poc-94} in data taking.

\end{document}